# Plants to Polyelectrolytes: Theophylline Polymers and their Microsphere Synthesis

Ryan Guterman,* Markus Antonietti, Jiayin Yuan*

__________

Department of Colloid Chemistry, Max Planck Institute of Colloids and Interfaces
Am Mühlenberg 1 OT Golm, D-14476 Potsdam, Germany.
Jiayin.Yuan@mpikg.mpg.de
__________

To extend fossil oil supplies, sustainable feed stocks for the production of useful reagents and polymers should be harnessed. In this regard, chemicals derived from plants are excellent candidates. While the vast majority of plant sources used for polymer science only contain $C_xH_yO_z$, alkaloids such as caffeine, nicotine, and theophylline possess nitrogen functionality that can provide new functions for bio-derived polymers and their synthesis. In this context, we exploited the chemistry of theophylline, a natural product found in chocolate and tea, to create a cationic "poly(theophylline)" in a straightforward fashion for the first time. We demonstrate how this new polymer can be synthesized and used for the creation of narrowly disperse cationic microspheres.

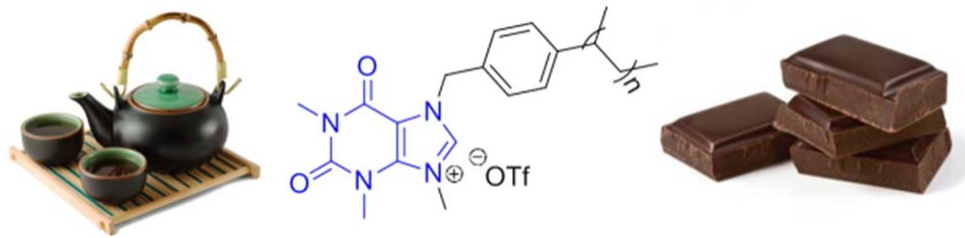

## 1. Introduction

The chemical industry has become reliant on the exploitation of petroleum to produce products for our modern society, even when many steps are required for the synthesis of functional and refined products. Researchers are encouraged to harness sustainable feed stocks for the production of useful reagents and polymers, especially for high value products with larger chemical distance to petroleum. In this regard, chemicals derived from plants are excellent candidates because they can be produced on large scale and possess functional group diversity. Some examples in this regard include phenols, terpenes, carbohydrates, and fatty acids, all of which have been used as sources for bio-derived materials.[1–6] To extend this bio-based approach, renewable resources must be manipulated in an efficient manner to meet the stringent requirements for application.[7–9] One class of compounds that is not commonly observed in polymer science includes alkaloids, which are a diverse class of nitrogen-containing organic molecules produced in plants. Some examples include nicotine, coniine, ephedrine, caffeine, and cocaine. Their natural production, amine functionality, complex hydrogen binding patterns, and in some cases, wide availability, makes their use for polymer and material science attractive. Alkaloids that are present in waste streams or in cultivated plants offer the greatest promise, most specifically nicotine and xanthines such as caffeine, theophylline, and theobromine. Theophylline, among other natural xanthines is easily extracted from plant material using selective solvents, such as dichloromethane, chloroform, and water.[10] The isolation of theophylline from cocoa is slightly different and may require a defattening step using a solvent such as petroleum ether prior to aqueous extraction.[11–13] Some examples where alkaloids have been used for materials science applications include nicotine-derived room temperature ionic liquids, which have been used for chiral recognition of carboxylic acids.[14] Caffeine has been used as a dopant for carbon-nitride,[15] flexible crystalline materials,[16,17] and recently as a monomer for free-radical polymerization by Long *et al.*, demonstrating the feasibility of this approach for the first time.[18] Harnessing alkaloids that possess many functional groups allows for multiple derivatization strategies and the production of polymers with additional function. Theophylline possesses N-H functionality (Figure 1A) which can be used for the attachment of a polymerizable group, and can also react with acids and electrophiles to produce cations, which provides easy entry into the field of functional polyelectrolytes.[19–21] Mecerreyes *et al.* previously demonstrated the formation of supramolecular ionic networks derived from partially[22] or completely natural sources,[23] however the fabrication of linear cationic polyelectrolytes from natural sources is still only weakly covered. Our interest in poly(ionic liquids), a subclass of polyelectrolytes comprised of ionic liquid monomer units, is derived from their use in novel sensors,[24] actuators,[25,26] and thermoresponsive materials,[27] thus demonstrating a potential application for alkaloid-containing polyelectrolytes. Theophylline is a commonly consumed alkaloid and is found naturally in cocoa beans and tea, and is currently used worldwide to treat asthma.[28] In this context, we are interested in covalent incorporation of this alkaloid into polymer chains and the investigation of their structure-property relationship and materials potential. We report the first synthesis of a theophylline-derived styrenic monomer, which was used to prepare a new cationic polyelectrolyte *via* radical polymerization and its colloidal particles *via* dispersion polymerization. The ability to form disperse microspheres provides the means to exploit these polyelectrolytes in new ways, as demonstrated across many fields, including catalysis,[29] sensing,[30,31] optics,[32] and biomedical technology,[33] where the synergistic properties of the polyelectrolyte anion/cation pair and the microsphere architecture can be harnessed.

## 2. Experimental Section

### 2.1 Materials

Theophylline (>99%), 4-vinylbenzyl chloride (>90%), and polyvinylpyrrolidone (PVP, 360 000 g/mol) were purchased from Sigma Aldrich and used as received. Azobisisobutyronitrile (AIBN, 98%) was purchased from Sigma Aldrich and recrystallized from methanol. Sodium hydride (NaH, 50% in mineral oil) was purchased from Alfa Aesar and rinsed with diethylether prior to use. Thermogravimetric analysis (TGA) experiments were conducted using a Netzsch TG209-F1 apparatus with a heating rate of 10 K min$^{-1}$ under nitrogen flow. Differential scanning calorimetry (DSC) experiments were performed on a Perkin-Elmer DSC-1 instrument at a heating/cooling rate of 5-20 K min$^{-1}$ under nitrogen flow. The melting points were determined from the 2$^{nd}$ heating curves. $^1$H, $^{13}$C{$^1$H}, and $^{19}$F{$^1$H} NMR spectra were collected on a Bruker DPX-400 spectrometer. Gel permeation chromatography (GPC) was performed using NOVEMA Max linear XL column with a mixture of 80% of aqueous acetate buffer and 20% of methanol. Conditions: flow rate 1.00 mL min$^{-1}$, PSS standards using RI detector Optilab DSP Interferometric Refractometer (Wyatt-Technology). Scanning electron microscopy (SEM) was performed using a Gemini LEO 1550 microscope at 3 kV. Mass spectrometry was performed on a Xevo G2-XS instrument by Waters for compound **1**, and on a Thermo Scientific Velos Pro instrument for all others. IR-ATR spectroscopy was performed on a Thermo Scientific iS5 instrument.

**2.2 Synthesis**

Synthesis of compound **1**.

Theophylline (5.0 g, 27.75 mmol) and 4-vinylbenzylchloride (8.47 g, 55.51 mmol) were added to 100 mL DMSO and cooled to 0 °C in an ice bath. NaH (2.66 g, 111.0 mmol) was then added slowly, and the mixture was kept cool for 30 minutes before warming to room temperature over 3 hours. The mixture was then slowly added to water (2.5 L) resulting in quenching of residual NaH and the formation of a precipitate, which was filtered off and rinsed with MeOH multiple times (200 mL in total). The white powder was dried *in vacuo* and identified as compound **1** (4.7 g, 57%). T$_m$ : 185 °C. $^1$H NMR (400 MHz, DMSO-$d_6$, δ) 8.27 (s, 1H), 7.43 (d, 2H, $^3J$ = 8 Hz), 7.30 (d, 2H, $^3J$ = 8 Hz), 6.85 (dd, 1H, (dd, 1H, $^3J$ = 18 Hz (trans), $^3J$ = 10 Hz (cis), C$H$=CH$_2$). 5.80 (dd, 1H, $^1J$ = 1 Hz, $^3J$ = 18 Hz), 5.25 (dd, 1H, $^1J$ = 1 Hz, $^3J$ = 18 Hz), 3.41 (s, 3H), 3.20 (s, 3H). $^{13}$C{$^1$H} NMR (100.5 MHz, DMSO-$d_6$, δ) 154.3,

150.9, 148.3, 142.5, 136.7, 136.4, 136.0, 127.8, 126.3, 114.6, 105.7, 48.7, 29.4, 27.5. FTIR-ATR: $v$ (cm$^{-1}$): 3103, 2975, 2949, 1710, 1655, 1604, 1544, 1513, 1471, 1456, 1403, 1369, 1327, 1289, 1229, 1191, 1125, 1026, 976, 905, 875, 836, 826, 806, 771, 762, 748, 718, 701, 639, 620, 566. HRMS m/z [M]$^+$ calc for C$_{16}$H$_{16}$N$_4$O$_2$ 296.1273, found 297.1348.

Synthesis of compound **2**.

Compound **1** (1.54 g, 5.36 mmol) was dissolved in 40 mL DCM followed by the slow addition of MeOTf (4.62 g, 28.15 mmol). The solution was stirred for 2.5 hours and then precipitated in 300 mL cyclohexane to remove excess MeOTf. The isolated powder was redissolved in acetone (60 mL) and precipitated in 600 mL Et$_2$O and dried *in vacuo*. A white powder was isolated and identified as compound **2** (1.95 g, 79%). T$_m$ : 155 °C. T$_{dec}$ = 305 °C. $^1$H NMR (400 MHz, DMSO-$d_6$, δ) 9.44 (s, 1H), 7.51 (d, 2H, $^3J$ = 8 Hz), 7.42 (d, 2H, $^3J$ = 8 Hz), 6.73 (dd, 1H, $^3J$ = 18 Hz (trans), $^3J$ = 11 Hz (cis), C*H*=CH$_2$) 5.87 (d, 1H, $^3J$ = 18 Hz), 5.69 (s, 2H), 5.30 (d, 1H, $^3J$ = 11 Hz) 4.15 (s, 3H), 3.72 (s, 3H), 3.26 (s, 3H). $^{13}$C{$^1$H} NMR (100.5 MHz, DMSO-$d_6$, δ) 153.0, 150.1, 139.8, 139.4, 137.6, 135.8, 133.4, 128.5, 126.5, 115.3, 106.9, 50.9, 37.0, 31.3, 28.4. $^{19}$F{$^1$H} NMR (376 MHz, DMSO-$d_6$, δ) -77.8. FTIR-ATR: $v$ (cm$^{-1}$): 3067, 1710, 1670, 1637, 1583, 1538, 1514, 1447, 1410, 1353, 1308, 1272, 1253, 1221, 1155, 1115, 1096, 1054, 1028, 1004, 930, 887, 856, 828, 780, 759, 746, 726, 668, 636, 603, 574. MS (ESI+): 311.2 ([Cation]$^+$), 771.3 ([(Cation)$_2$ + (Anion)]$^+$). MS (ESI-): 609.1 ([(Cation) + (Anion)$_2$]$^-$).

Synthesis of polymer **3**.

Compound **2** (0.776 g, 1.687 mmol) and AIBN (0.0155 g, 0.095 mmol) were dissolved in 1.3 mL DMSO, purged with N$_2$ gas for 10 minutes, and then heated to 65 °C for 16 hours. The viscous mixture was diluted with 25 mL MeCN and precipitated in THF (250 mL) three times, and dried *in vacuo* at 60 °C. The white powder was identified as polymer **3** (0.71 g, 91%). Mn = 1.54 x 10$^5$ g/mol, Đ = 3.8. T$_{dec}$ = 305 °C. $^1$H NMR (400 MHz, MeCN-$d_3$, δ) 8.83 (s, 1H), 7.14-6.54 (m, 4H), 5.61 (s, 2H), 4.13 (bs, 3H), 3.74 (bs, 3H), 3.07 (bs, 3H), 1.43 (bs,

2H). $^{19}$F{$^1$H} NMR (376 MHz, MeCN-$d_3$, δ) -79.206. FTIR-ATR: $v$ (cm$^{-1}$): 3600, 3150, 2900, 1720, 1668, 1634, 1575, 1538, 1461, 1348, 1253, 1224, 1155, 1050, 1028, 1003, 818, 779, 770, 759, 745, 636, 604, 574.

Dispersion Polymerization

Microspheres synthesis was conducted according to a modified procedure.[34] Briefly, **2** (0.1 g, 0.217 mmol), PVP (5 mg), and AIBN (4 mg) were mixed with the methanol/ethanol solvent (1 mL). After cooling to 0 °C the solutions were purged with N$_2$ gas for 10 minutes and then heated to 70 °C for 6-8 hours. The microspheres were rinsed with the reaction solvent and dropcast on glass slides for SEM analysis.

## 3. Results and Discussion

The key monomer was synthesized by first treating theophylline with 4-vinylbenzylchloride under basic conditions in DMSO (Figure 1A). Quenching of the residual NaH and precipitation of the hydrophobic product occurred simultaneously upon addition of the mixture to water, followed by a MeOH rinse to isolate the pure product **1**. We initially tried to quaternize **1** with methyl iodide, as reported for similar compounds like caffeine.[35,36] However the harsh reaction conditions required (>70 °C, 24 hrs) resulted in undesired side reactions (Figure S4). Further attempts to quaternize **1** using methyl iodide under a variety of conditions were unsuccessful. Instead, we believed that a stronger methylating agent that can be used at room temperature would be more promising. Using excess methyl triflate, *N*-methylation of **1** in DCM was complete after three hours at room temperature, and the product was precipitated in cyclohexane and diethyl ether to isolate **2** (Figure 1B). In the $^1$H NMR spectra in Figure 1B, a downfield shift was observed for the RN(C*H*)NR proton (8.27 to 9.44 ppm), consistent with the quaternization of caffeine.[12] We found that extended reaction times even with lower amounts of methyl triflate led to parallel decomposition of the product (Figure S5), while shorter reaction times in the presence of excess methyl triflate led to cleaner conversion. The melting point was reduced from 185 °C for **1** to 155 °C for **2**, indicating that the large size and conformational flexibility of the ions resulted in a reduction of its melting point. While the synthesis of **2** requires two derivitization steps including the use of a strong base and alkylating agent, we find that other cationic monomers used currently require similar derivitization processes.[37–39] By using naturally sourced amines we demonstrate a greener alternative to the synthesis of cationic monomers.

This monomer was polymerized by conventional free-radical polymerization followed by precipitation in THF three times to remove residual monomer and isolate polymer **3** in good yield (91 %). Broadening of the aryl and methyl proton signals was observed along with the introduction of new signals at δ = 2.98 and 0.50-1.50 (Figure S9), which are attributed to the polymer backbone protons. GPC analysis of **3** revealed a polydisperse sample (Đ =3.8) with an apparent number-average molecular weight of 1.54 x $10^5$ g/mol. This polymer possessed no discernable glass transition up to a temperature of 250 °C by DSC analysis. Compared to its monomer, both **2** and **3** are thermally stable up to 300 °C.

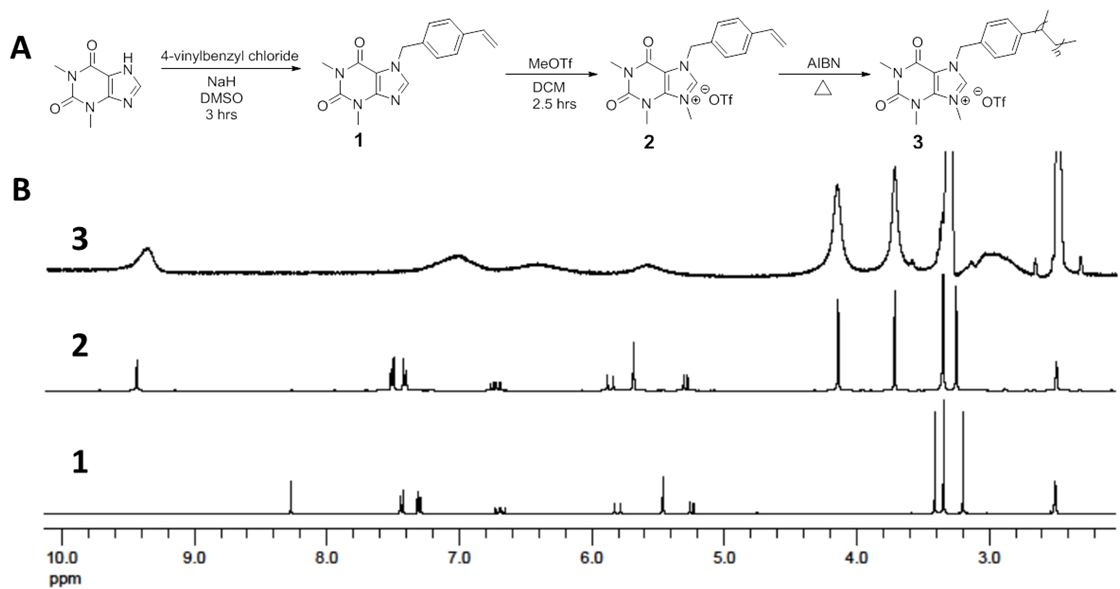

**Figure 1:** A) Synthetic scheme of compounds **1-3**. B) $^1$H NMR spectrum of **1-3** in DMSO-$d_6$.

Table S1: Solubility of **1-3** in various solvents*

| Compound # | H₂O | MeOH | Acetone | THF | MeCN | DMSO | CHCl₃ | Cyclohexane |
|---|---|---|---|---|---|---|---|---|
| 1 | Ins | Sol | Sol | Sol | Sol | Sol | Sol | Ins |
| 2 | Ins | Sol | Sol | Sol | Sol | Sol | Sol | Ins |
| 3 | Ins | Ins | Ins | Ins | Sol | Sol | Ins | Ins |

*Solubility determined at a concentration of 8 mg/mL.

The solubility properties of **1-3** were also elucidated (Table 1). Both **1** and **2** were soluble in all polar organic solvents tested (*i.e.* MeOH, acetone, THF). However **3** was found to be insoluble in all solvents except in DMSO and acetonitrile. This dramatic change in solubility upon polymerization is consistent with the loss of dissolution entropy by polymerization, i.e. polymers dissolve in fewer solvents than their monomers. The solubility difference of monomer and polymer in the same solvent can be utilized for the synthesis of uniform microspheres by dispersion polymerization.[34] In this method, the sphere size can be tuned by adjusting the polarity of the reaction solvent with a co-solvent, where low polarity media favours smaller sphere sizes. The use of alcohols for this purpose is convenient as higher alcohols are progressively more unipolar and are worse solvents for ionic molecules.[40,41] In our case, **2** was mixed with PVP (5 wt%) and AIBN (4 wt%) in different methanol/ethanol mixtures (0, 25, 50, and 80 wt% ethanol in methanol) and heated to 70 °C for 6-8 hours in a sealed flask. In pure methanol, microspheres varied greatly in size and were overall of poor quality in terms of the polydispersity (Figure 2A). However with 25% ethanol addition, a bimodal distribution was observed with sphere sizes centered at 0.9 and 1.8 microns in diameter, respectively (Figure 2B, and Figure 3A). Narrowly dispersed polymer spheres were produced using 50% ethanol, with an average size of 1.43 ± 0.1 microns (Figure 2C), while 80% ethanol decreased the size of the spheres further to 0.59 ± 0.1 microns (Figure 2D). This trend is consistent with similar dispersion polymerizations, where the introduction of an antisolvent drives particle nucleation, thus leading to smaller spheres. In our system however, the introduction of antisolvent not only decreases sphere size, but promotes a monomodal distribution.

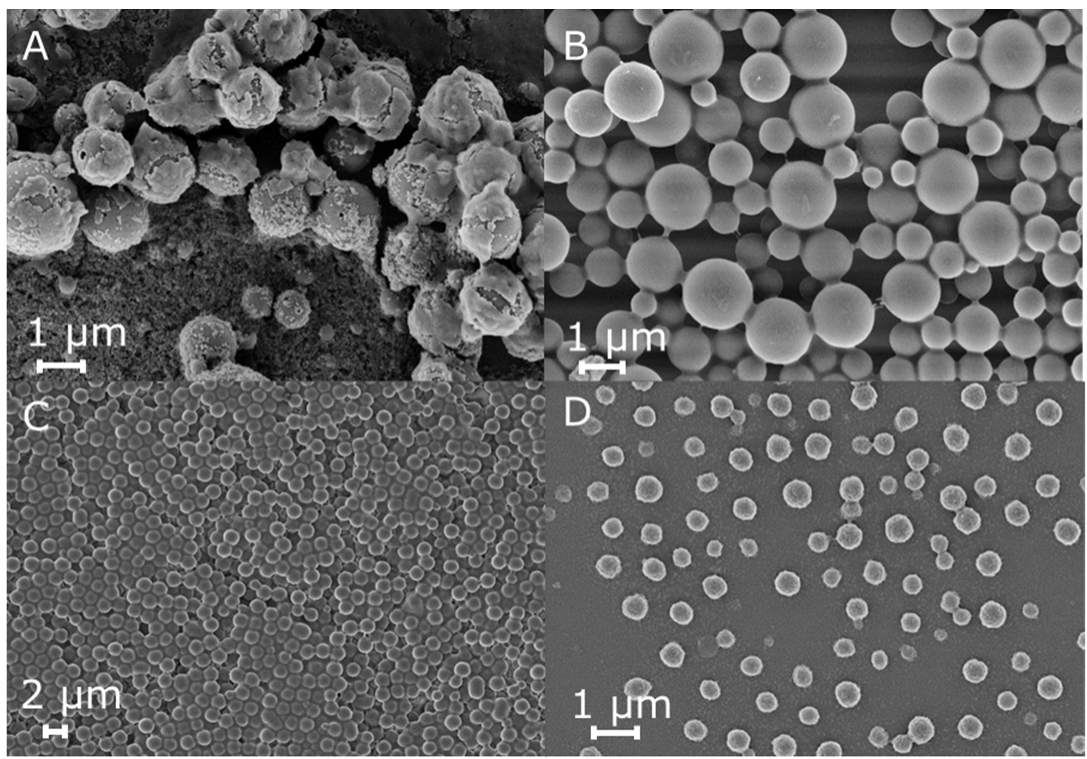

**Figure 2:** SEM images of microspheres prepared by dispersion polymerization in different methanol:ethanol mixtures. A) 100:0 B) 75:25 C) 50:50 and D) 20:80.

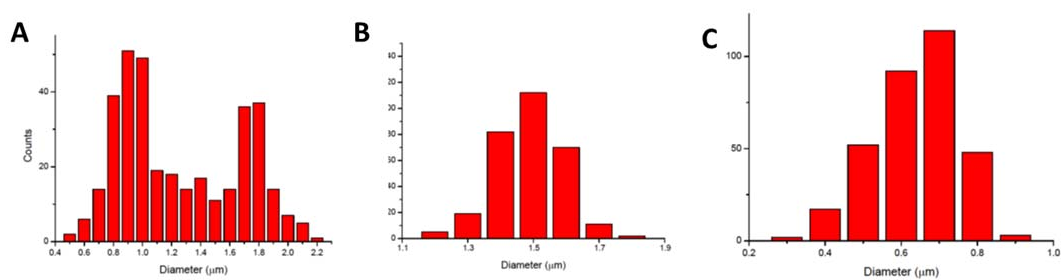

**Figure 3**: Particle size distributions for microspheres prepared in different methanol:ethanol mixtures of 75:25 (A), 50:50 (B), and 20:80 (C).

## 4. Conclusions

The synthesis of *N*-functionalized theophylline with a polymerizable group was accomplished and converted in to a cationic salt using methyl triflate in a simple fashion. This monomer was polymerized by free radical polymerization to create a cationic polyelectrolyte. *Via* dispersion polymerization, narrowly dispersed alkaloid-containing polyelectrolyte microspheres were prepared whose polydispersity and size could be altered by tuning the reaction medium. Our approach for cationic polyelectrolyte synthesis using naturally found alkaloids is attractive from a sustainable viewpoint and represents our latest and continuing efforts to prepare functional polyelectrolytes from plant-based materials as feedstock sources. While theophylline is not the most abundant alkaloid in nature, the methodology presented in this work can be applied to more abundant alkaloids.

**Supporting Information**

Supporting Information is available from the Wiley Online Library or from the author.

Acknowledgements: This work was supported by the European Research Council Starting Grant (639720-NAPOLI).

# Plants to Polyelectrolytes: Theophylline Polymers and their Microsphere Synthesis

Ryan Guterman,* Markus Antonietti, Jiayin Yuan*
*Department of Colloid Chemistry, Max Planck Institute of Colloids and Interfaces*
*Am Mühlenberg 1 OT Golm, D-14476 Potsdam, Germany*
E-Mail: jiayin.yuan@mpikg.mpg.de

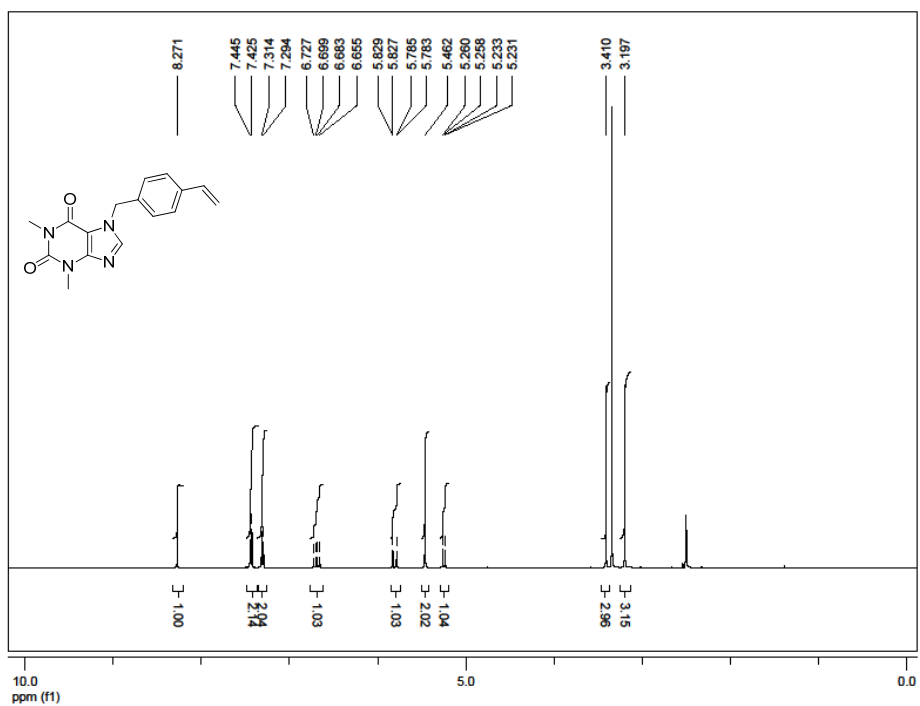

**Figure S1:** $^1$H NMR spectrum of **1** in deuterated DMSO.

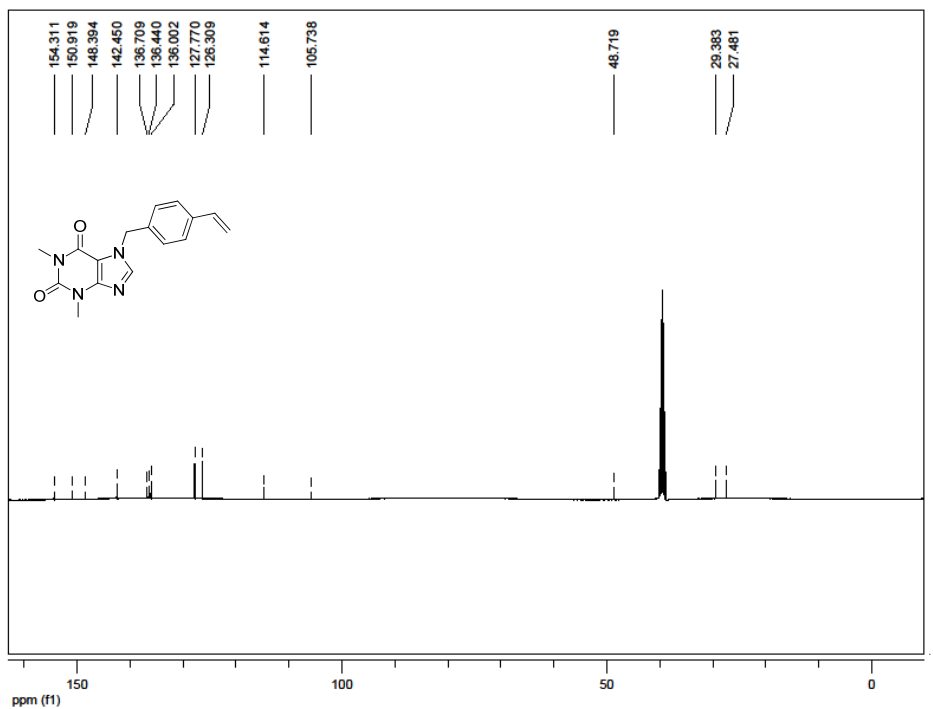

**Figure S2:** $^{13}C\{^1H\}$ NMR spectrum of **1** in deuterated DMSO.

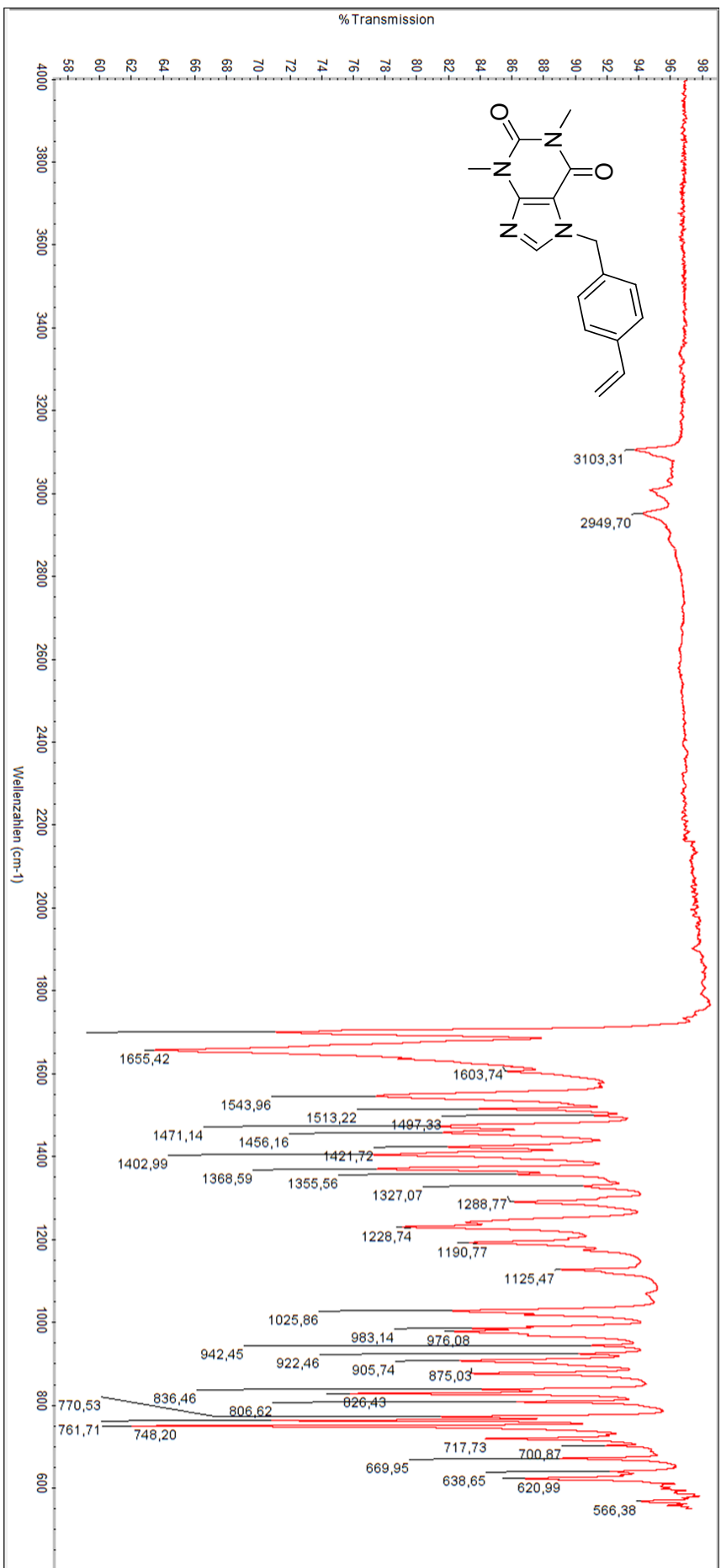

**Figure S3:** ATR-IR spectrum of **1**.

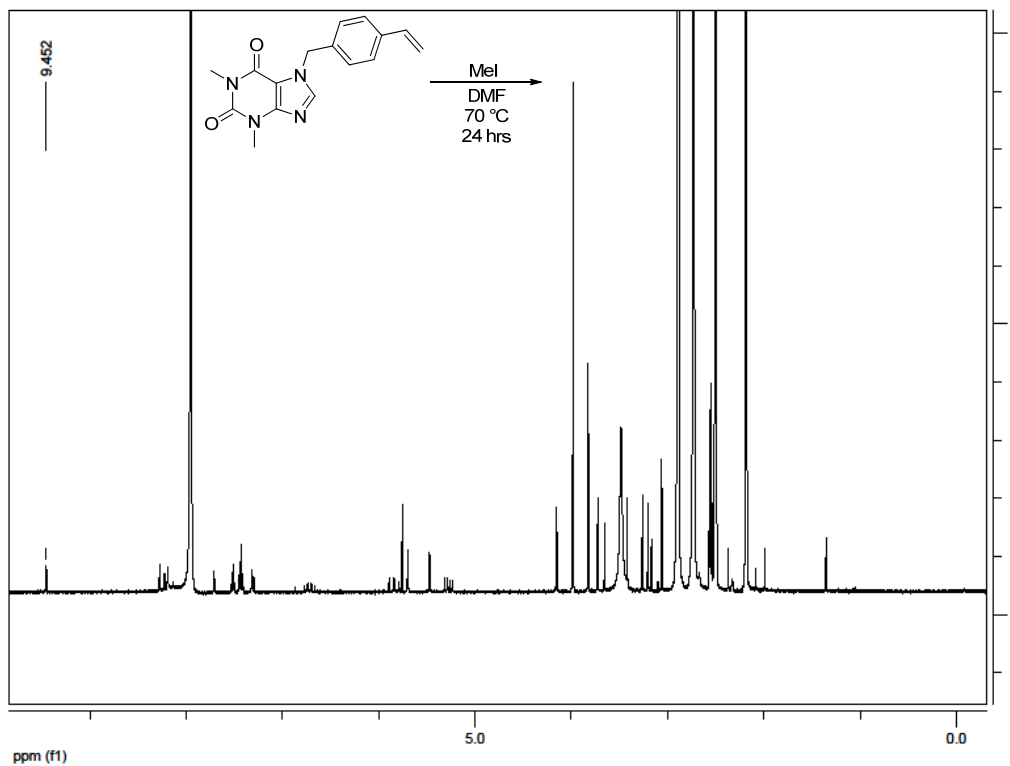

**Figure S4:** $^1$H NMR spectrum of **1** (0.2 g, 0.6752 mmol) in 10 mL of a 3:1 mixture of DMF and MeI at 70 °C for 24 hrs. Experiment performed in deuterated DMSO.

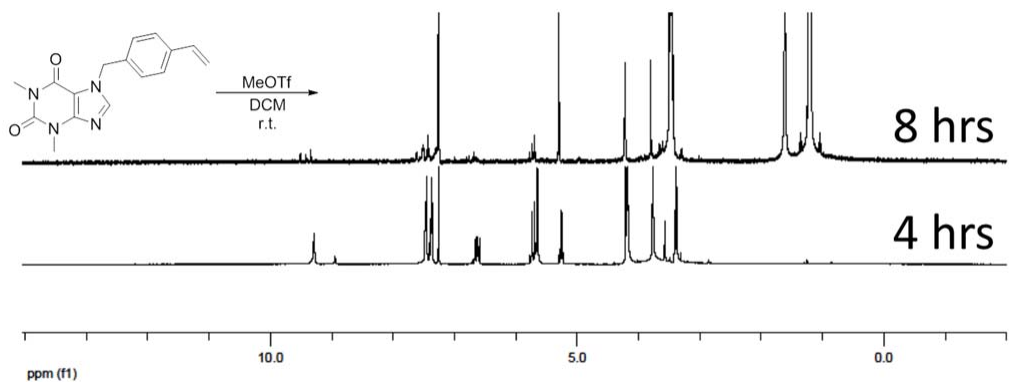

**Figure S5:** $^1$H NMR spectra of **1** (1.40 g, 4.872 mmol) with methyl triflate (2.50 g, 14.616 mmol) in 10 mL DCM. The reaction was mostly complete after 4 hrs, however after 8 hours significant decomposition was observed. Experiment performed in deuterated chloroform.

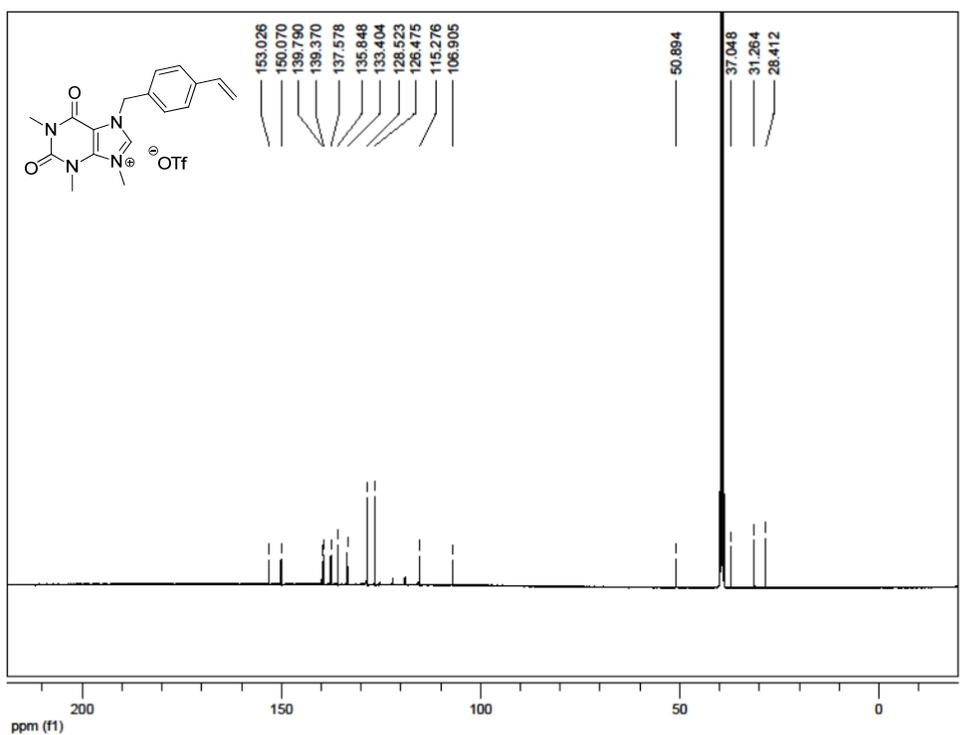

**Figure S6:** $^{13}$C{$^1$H} NMR spectrum of **2** in deuterated DMSO.

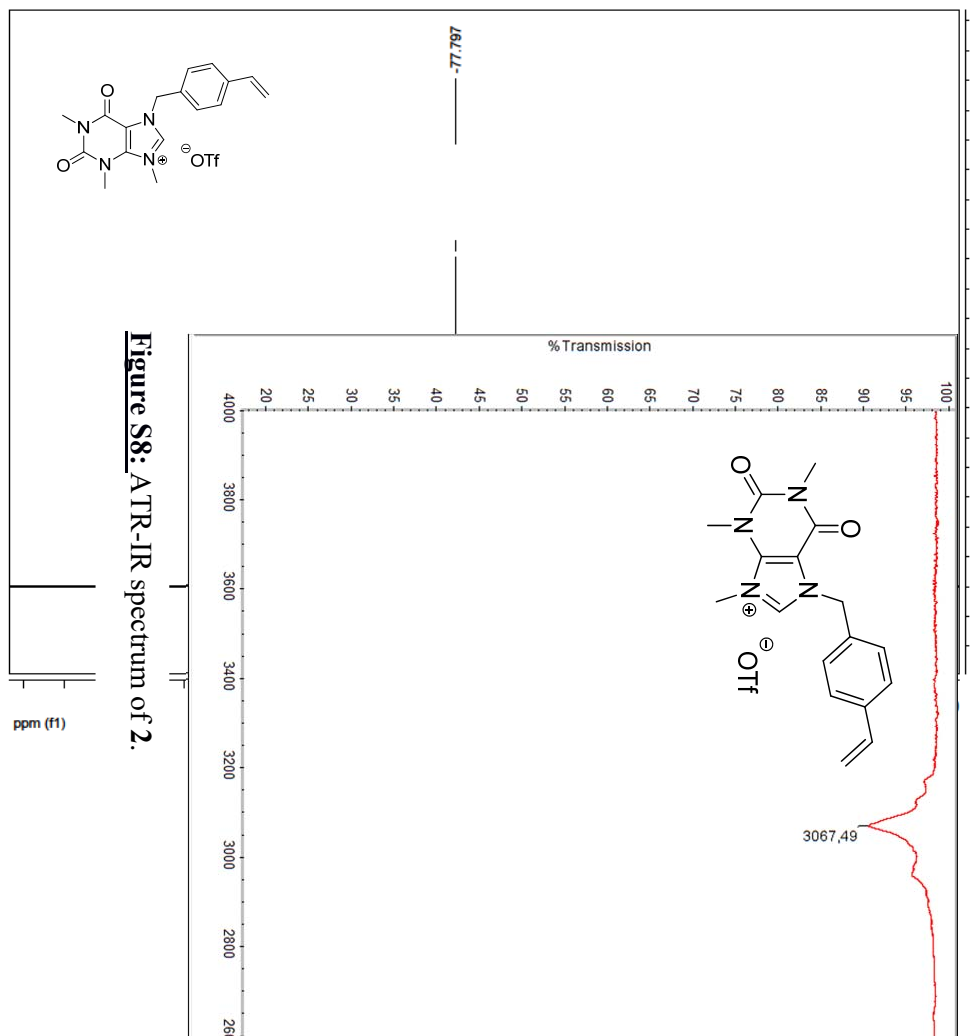

**Figure S8:** ATR-IR spectrum of **2**.

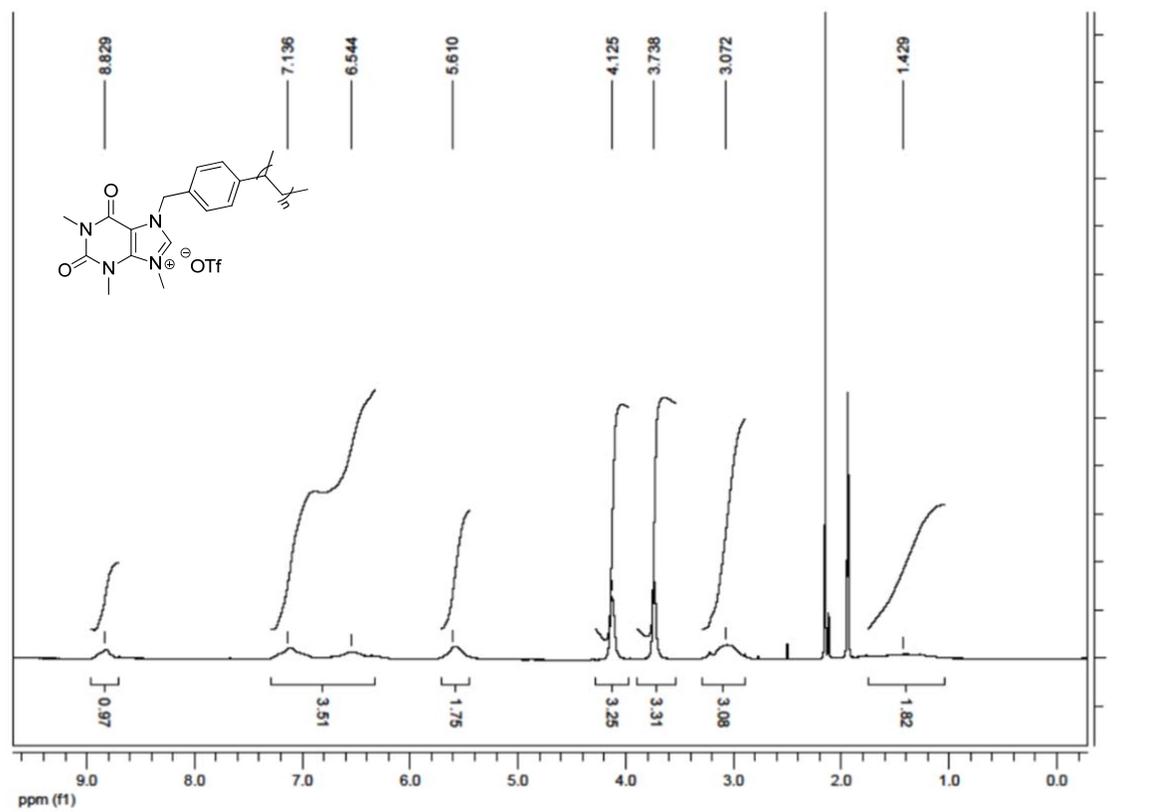

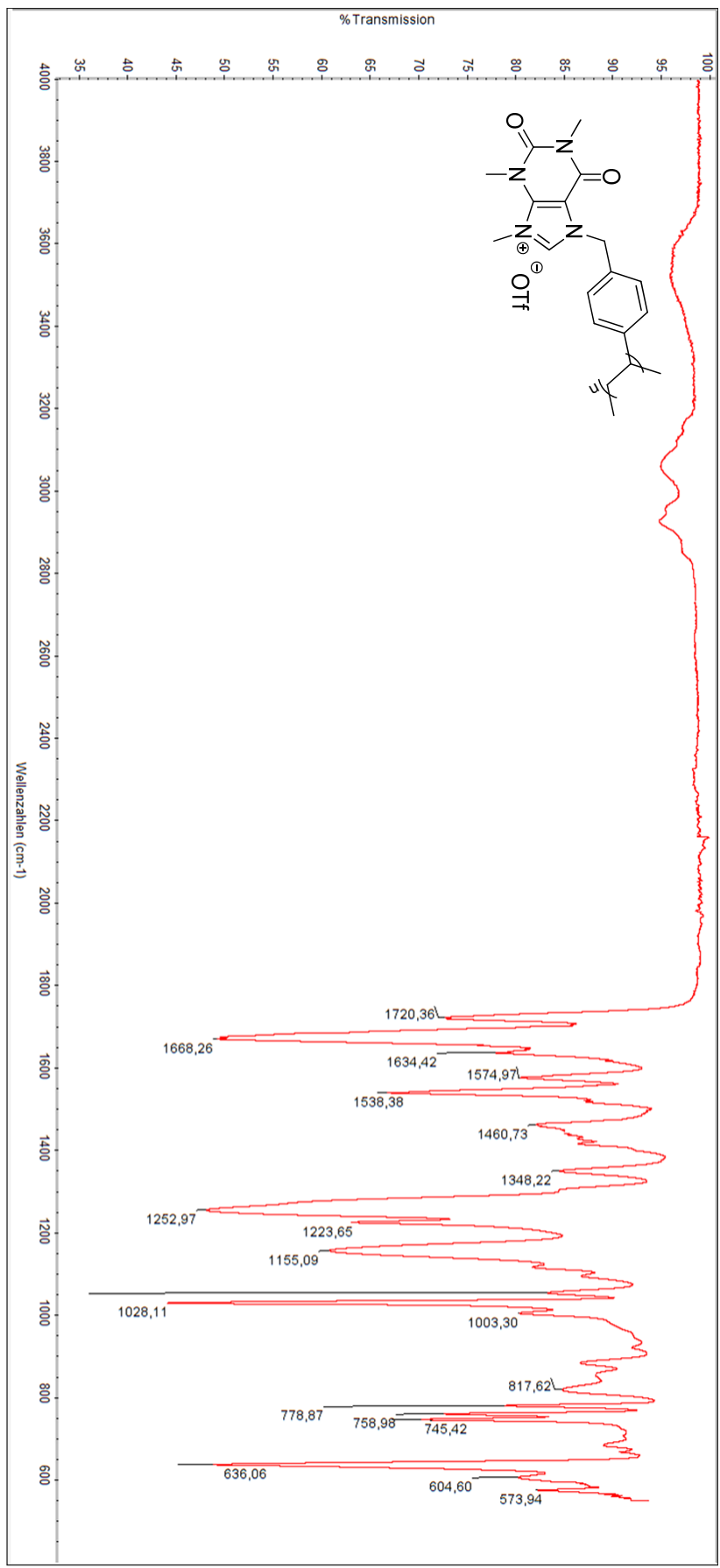

**Figure S10:** ATR-IR spectrum of **3**.

**Figure S11:** TGA plots of **2** and **3** at 10 K/min heating rate under a nitrogen atmosphere.

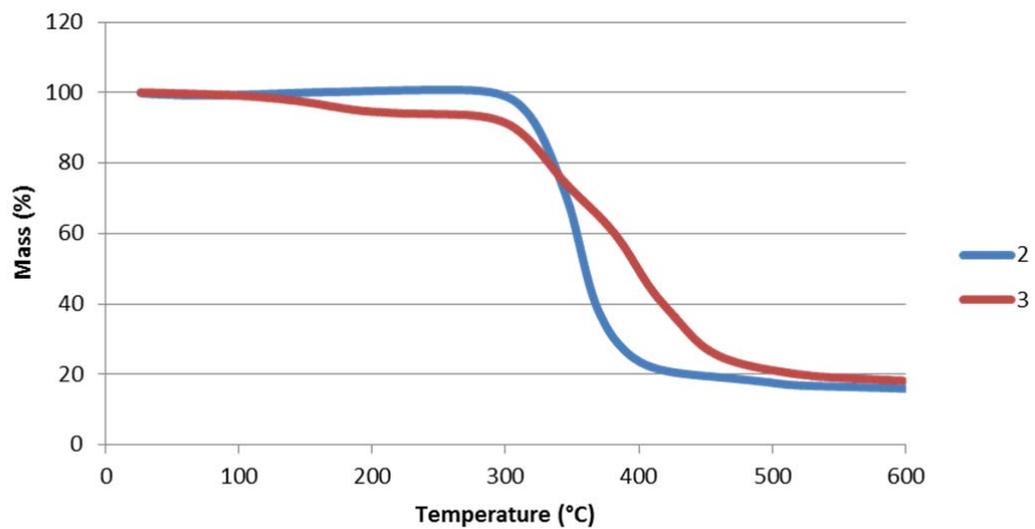